\documentstyle[12pt]{article}
\date{}
\def\a{\alpha}\def\b{\beta}\def\d{\delta}

\def\g{\gamma}

\def\l{\lambda}
\def\r{\rho}\def\s{\sigma}

\def\L{\Lambda}
\def\O{\Omega}

\def\st{spacetime }
\def\fe{field equations }\def\as{asymptotically }
\def\tran{transformations }\def\coo{coordinates }

\def\rep{representation}
\def\cf{conformally flat }
\def\cur{curvature }\def\ms{maximally symmetric
}
\def\coot{coordinate transformation }
\def\ads{anti-de
Sitter }

\def\des{de Sitter
}

\def\et{e^{2t}}\def\ab{{\a\b}}\def\AB{{AB}}
\def\kc{constant curvature }
\def\dis{d\s_k^2}\def\ak{a_k^2(t)}

\def\section#1{\bigskip\noindent{\bf#1}\smallskip}

\def\ref#1{\medskip\everypar={\hangindent 2\parindent}#1}
\def\beginref{\begingroup
\bigskip
\centerline{\bf References}
\nobreak\noindent}
\def\endref{\par\endgroup}

\begin{document}

\baselineskip18pt

{
\centerline{\large 
{\bf Classification of multidimensional inflationary
models}}
\vskip50pt
\centerline{\bf S. Mignemi$^*$ and H.-J. Schmidt$^\dagger$}
\vskip20pt
\centerline{$^*$ Dipartimento di Matematica, Universit\`a di
Cagliari}
\centerline{viale Merello 92, I-09123 Cagliari, Italy}
\smallskip
\centerline{$^\dagger$ Universit\"at Potsdam, Institut f\"ur
Mathematik}
\centerline{D-14415 Potsdam, PF 601553, Am Neuen Palais 10,
Germany}
\vskip70pt
\centerline{\bf Abstract}
\bigskip
\noindent We define under which circumstances two multi-warped
product
spacetimes can be considered equivalent and then we classify the
spaces of
constant curvature in the Euclidean and Lorentzian signature. For
dimension
$D=2$, we get essentially twelve representations, for $D=3$
exactly eighteen. More general,  
for every even $D$, $5D+2$ cases exist, whereas for every 
odd $D$, $5D+3$  cases exist. For every $D$, exactly
one half of them has the Euclidean signature. 
 Our definition is well
suited for
the discussion of multidimensional cosmological models, and our
results give 
a simple algorithm to decide whether a given metric represents
the
inflationary de Sitter spacetime (in unusual coordinates) or not.
\vskip90pt
\noindent 
J. Math. Phys. submitted \# 7 - 343 

of cosmology; 
  0240.Ky Riemannian and Pseudoriemannian geometries  
\vfil\eject}

\section{I. Introduction.}

Recently, a lot of papers dealt with multi-dimensional
cosmological models,
cf. e.g. [1,2,3] and references cited there. In these works, the
ansatz for
the metric is a generalized warped product of several spaces,
with the warping
functions depending on the time $t$ only, i.e.
$$ds^2=dt^2-\sum_{k=1}^n\ak\ \dis\eqno(1.1)$$
where each $\dis$ represents a Riemannian space of dimension
$d_k\ge1$.
In the usual interpretation, one requires $d_1=3$ and $d\s_1^2$
to be the
physical 3-space, whereas all other spaces $d\s_k^2$, $k\ge2$ are
internal
spaces. However, our approach is not restricted to this
interpretation.

One of the often discussed questions in this context is the
appearence of
an inflationary phase of cosmic evolution. Its geometry is
usually
represented by a \st of constant negative curvature, called \des
spacetime.
It is well-known how to represent the \des \st of arbitrary
dimension in
the form (1.1), but up to now there does not 
exist a classification of
the 
metrics of the general form (1.1) to decide 
which of them correspond to spaces of
constant curvature.

To elucidate what this means, let us consider the metrics (1.2)
and
(1.3).
$$ds^2=dt^2-\sinh^2 t\ dx^2-\cosh^2 t\ (dy^2+\sin^2y\
dz^2)\eqno(1.2)$$
(see e.g. [4, eq.(3.18)]). The metric (1.2) is of the type (1.1)
and is
obviously a cosmological model of Kantowski-Sachs type. The
metric (1.2) is
known to represent a vacuum solution of Einstein's equations with
a positive
cosmological term. In agreement with the cosmological no-hair
theorem it is
\as de Sitter. In the \coo used, the metric (1.2) seems to define
an
anisotropic model, however (cf. [4]) it is nothing but a piece of
the
isotropic \des\st in unusual coordinates, and moreover, the \coot
necessary to
bring it to the classical form is quite involved.

A related metric is (cf. [5, eq. (4.6)]),
$$ds^2=dt^2+\sin^2t\ dx^2+\cos^2t\ dy^2\eqno(1.3)$$
which is the metric of the standard 3-sphere in unusual
coordinates.

It is the purpose of the present paper to give a complete
classification of
all the metrics of type (1.1) which represent spaces of constant
curvature.

The paper is organized as follows: sect. II introduces the
generalized warped
products and defines the normal form of the metric (1.1). Sect. 
III gives the
results for the dimension $D\le 3$ and sect. IV covers the
remaining cases
$D\ge4$. In sect. V, we consider a slightly different 
form of the 
metric,
where the $a_k$ can additionally 
depend on one of the 
 spatial coordinates, but the $d\s_k^2$
have
dimension 1. Sect. VI discusses the results.

\bigskip

\section {II. Generalized warped products.}

Usually, a warped product is the conformally transformed
cartesian product
between two Riemannian spaces, where the conformal factor,
called warping
function, depends on the \coo of one of the factors only. If
there are more
than two factor spaces, then there exist several possible
generalizations of
that notion. We will restrict ourselves to the following types of
generalized
warped products: all $a_k>0$ and
$$ds^2=dt^2\pm\sum_{k=1}^n\ak\ \dis\eqno(2.1)$$
and each $\dis$ is a Riemannian space of dimension $d_k\ge1$.
Thus we
consider the Euclidean (upper sign) and the Lorentzian (lower
sign) signatures
of the metric. In this way we are able to cover both eqs. (1.2)
and (1.3).
 Obviously, the metric (2.1) represents a usual warped 
 product for
 $n=1$.
The dimension of $ds^2$ is $D=1+\sum d_k$. We want to find out
under which
circumstances the metric (2.1) represents a space of constant
curvature
locally (so we do not discuss whether zeroes of the functions
$a_k(t)$ give
rise to coordinate singularities and whether our metrics are
geodesically
complete).

To get a uniquely defined classification, we start by explaining
under which
circumstances two metrics of type (2.1) will be considered to be
the same:

A) Any constant $c_k>0$ can be shifted between $a_k^2(t)$ and
$d\s_k^2$;

B) $ds^2$ can be multiplied by a positive constant, or in other
words: $t$ and
every $a_k(t)$ can be multiplied by the same positive factor;

C) Linear \tran of $t$ will be allowed;

D) Permutations of the indices $k$ are allowed;

E) Coordinate \tran involving only the \coo of one of the spaces
$d\s_k^2$ are
allowed.

F) If two warping functions coincide, $a_j(t)=a_k(t)$, 
($j\ne k$), 

\noindent 
then
$a_j^2(t)d\s_j^2+a_k^2(t)d\s_k^2$ can be replaced by
$a_j^2(t)(d\s_j^2+
d\s_k^2)$.

Comment: B) means that we do not distinguish between
homothetically equivalent
spaces, or, in other words, a change of the length unit will not
be considered
essential. F) means that the number $n$ of terms in the sum of
eq. (2.1) is
not fixed. However, we can define the normal form of the
representation (2.1)
as follows: whenever possible (if necessary by the help of
introducing suitable
constants $c_k$ according to A), the replacement F) shall be done
(of course,
this procedure terminates uniquely). Equivalently we may say: the
normal form of
the representation is that one where the number $n$ is minimal.
Furthermore it holds: the metric (2.1) is in normal form if the
quotients
$a_j(t)/a_k(t)$ are constant for $k=j$ only.

For technical reasons, we will sometimes discuss spaces in
non-normal form.
However, if the number $n$ of factors is explicitly mentioned,
we always
refer this to the normal form and, without loss of generality, we
may restrict
to metrics (2.1) in normal form.

\bigskip

\section {III. Classification for dimension $D\le 3$.}

$D=1$ implies $n=0$ and $ds^2=dt^2$ represents flat space. So we
have just
one representation for $D=1$.

\medskip

\section {III.1. Dimension $D=2$.}

Let $D=2$, then $d\s_1^2=dx^2$ is flat\footnote{Spaces
of \kc in $D=2$ have been discussed, for example, in [6].}.      

    We write $a(t)$ instead of $a_1(t)$ and get
$$ds^2=dt^2\pm a^2(t)\ dx^2\eqno(3.1)$$
which is already in normal form. This space is of constant
curvature iff
the curvature scalar $R=-2\ddot a/a$ is constant, where a dot
denotes $d/dt$.
Both signatures can be considered simultaneously, because they
are related
by the imaginary \coot $x\to ix$. (Here we give the derivation
for the known case
$D=2$ in detail in order to show how our formulation works, we do
not
repeat the analogous arguments in the other cases). According to
B), we may
restrict to the three cases $R=0$, $R=2$ and $R=-2$. For $R=0$,
$a(t)$ must
be linear in $t$, and according to A) and C), the following
representations
are singled out:
$$ds^2=dt^2\pm dx^2\eqno(3.2)$$
and
$$ds^2=dt^2\pm t^2dx^2\eqno(3.3)$$
Clearly, these are the cartesian and polar \coo respectively for
the flat
space. For $R=2$, $\ddot a+a=0$ has to be solved. According to A)
and C), only
one representation, the standard metric of the 2-sphere, appears:
$$ds^2=dt^2\pm \sin^2t\ dx^2\eqno(3.4)$$
For $R=-2$, however, three different representations appear for
each of the two
signatures:\footnote{By condition C) the solution
with $e^{-2t}$ instead of $\et$
can be excluded.}
$$ds^2=dt^2\pm\et\ dx^2\eqno(3.5)$$
further
$$ds^2=dt^2\pm\sinh^2t\ dx^2\eqno(3.6)$$
and
$$ds^2=dt^2\pm\cosh^2t\ dx^2\eqno(3.7)$$
An explicit \coot between the metrics (3.5) and (3.6) is deduced
in the appendix
of ref. [4]. Metric (3.4) and (3.6) are related by multiplying
$s$, $t$ and $x$
by $i$.

Thus we have exactly twelve different \rep{}s
 of a space of constant
curvature for
$D=2$. Six of them have $R<0$ and only two have $R>0$.

\medskip

\section {III.2.  Dimension $D=3$. First part.}

Let $D=3$. The space is of constant curvature iff the three
eigenvalues of the
Ricci tensor coincide. Let us start recalling the known cases.
The cartesian
product of flat spaces is flat. So we can get from eq. (3.2),
(3.3) for 
the
$R=0$ cases:
$$n=1,\qquad\qquad ds^2=dt^2\pm
(dx^2+dy^2)\eqno(3.8)$$
and  
$$n=2,\qquad\qquad ds^2=dt^2\pm (t^2\ dx^2+dy^2)\eqno(3.9)$$
For $R>0$ we know the examples
$$n=1,\qquad\qquad ds^2=dt^2+\sin^2t\ (dx^2+\sin^2x\
dy^2)\eqno(3.10)$$
and
$$n=2,\qquad\qquad ds^2=dt^2\pm (\sin^2t\ dx^2+\cos^2t\
dy^2)\eqno(3.11)$$
Here, (3.10) is the standard three-sphere and (3.11) is the
metric (1.3) [5].
Replacing $x$ by $ix$, we get from the metric (3.10) the
Lorentzian signature
with $R>0$:
$$n=1,\qquad\qquad ds^2=dt^2-\sin^2t\ (dx^2+\sinh^2x\
dy^2)\eqno(3.12)$$
For $R<0$ we get analogously, as in sect. III.1:
$$n=1,\qquad\qquad ds^2=dt^2\pm\et\
(dx^2+dy^2)\eqno(3.13)$$
and
 $$n=2,\qquad\qquad ds^2=dt^2\pm (\sinh^2t\ dx^2+\cosh^2t\
dy^2)\eqno(3.14)$$
(cf. [5, eq.(3.9)]). So, we have again twelve different cases.
However,
before checking their completeness in section III.4, we have to
discuss
the case $n=1$ in the next section.

\medskip

\section {III.3. The case $n=1$ in arbitrary dimension.}

Let $D\ge3$ be arbitrary and be $n=1$ in eq.(2.1). So the index 1
may be omitted
and $d=D-1$. We write $d\s^2=h_\ab dx^\a dx^\b$, where  $h_\ab$
depends on the
spatial \coo only, $\a,\b=1,...,d$. For $ds^2=dt^2\pm
a^2(t)d\s^2$ we get
$R_{00}=-d\ddot a/a$. Indices $j,k,...$ take the $D$ values
$0,1,...,d$.
If $ds^2$ is a space of constant
curvature, then it must be an Einstein space, i.e. Ricci tensor
and metric
tensor differ by a constant factor only. So we may assume,
without loss of
generality due to condition B), that
$$R_{jk}=\l dg_{jk},\qquad\quad {\rm with}\
\l\in\{-1,1,0\}\eqno(3.15)$$
Now, $g_{00}=1$ and we set $\l=-\ddot a/a$. So we are in a case
similar to
that of sect. III.1. Consequently, only the same six functions
$a(t)$ may appear:
$a=1$ and $a=t$ for $R=0$, $a=\sin t$ for $R>0$, and $a=e^t$,
$a=\sinh t$,
$a=\cosh t$ for $R<0$.

Let us now apply the $\a\b$-component of eq. (3.15), and let
$P_\ab$ denote
the Ricci tensor of $d\s^2$. We get:
$$P_\ab=R_\ab\pm h_\ab[a\ddot a+(d-1)\dot a^2]\eqno(3.16)$$
From (3.15) and (3.16) we get
$$P_\ab=\L h_\ab,\qquad\qquad\L=\pm(d-1)(\dot a^2+\l
a^2)\eqno(3.17)$$
It is easy to see that $\L$ is a constant. Concluding, for $n=1$,
$d\s^2$
must be an Einstein space of constant curvature scalar.

\medskip

\section{III.4. Dimension $D=3$. Second part.}

Now we are ready to complete the discussion for $D=3$. For $n=1$,
we have
$d=2$, and by sect. III.3, $d\s^2$ has constant curvature scalar.
This means
that $d\s^2$ is of constant curvature here, and (3.17) reads
$\dot a^2+\l a^2=
\pm\L$. $\l=0$, $a=1$ gives the metric (3.8). $\l=0$, $a=t$,
gives $\L=\pm1$,
namely
$$n=1,\qquad\qquad ds^2=dt^2+t^2 (dx^2+\sin^2x\
dy^2)\eqno(3.18)$$
which is flat space in spherical \coo, and
$$n=1,\qquad\qquad ds^2=dt^2-t^2 (dx^2+\sinh^2x\
dy^2)\eqno(3.19)$$

$\l=1$, $a=\sin t$ gives $\L=\pm1$, i.e. the metric (3.10) and
(3.12).
$\l=-1$, $a=e^t$ yields $\L=0$, i.e. the metric (3.13).
$\l=-1$, $a=\sinh t$ yields $\L=\pm1$, i.e.,
$$n=1,\qquad\qquad ds^2=dt^2+\sinh^2t\ (dx^2+\sin^2x\
dy^2)\eqno(3.20)$$
and
 $$n=1,\qquad\qquad ds^2=dt^2-\sinh^2t\ (dx^2+\sinh^2x\
dy^2)\eqno(3.21)$$
Finally, $\l=-1$, $a=\cosh t$ yields $\L= \mp 1$, i.e. 
$$n=1,\qquad\qquad ds^2=dt^2+\cosh^2t\
(dx^2+\sinh^2x\ dy^2)\eqno(3.22)$$
and 
$$n=1,\qquad\qquad ds^2=dt^2-\cosh^2t\ (dx^2+\sin^2x\
dy^2)\eqno(3.23)$$

This completes the discussion for $n=1$, and the six metrics
(3.18)-(3.23)
together with the twelve metrics of section III.2 form already
eighteen different
\rep{}s of
 a space of \kc in $D=3$. It remains to check completeness
for the
case $n=2$. So we have to use the ansatz
$$ds^2=dt^2\pm(a^2(t)\ dx^2+b^2(t)\ dy^2)\eqno(3.24)$$
and have to ensure that the three eigenvalues of the Ricci tensor
of (3.24)
coincide. This is equivalent to
$${\ddot a\over a}={\ddot b\over b}={\dot a\over a}{\dot b\over
b}
\eqno(3.25)$$
A detailed investigation of eq. (3.25) shows that all solutions
are already
covered by the eighteen cases mentioned before. 

Let us summarize the results for $D=3$: six representations with
$n=2$ and
twelve representations 
 with $n=1$ form a complete classification. Six of
these are flat,
four have positive curvature and the remaining eight have $R<0$.

\bigskip

\section{IV. Classification for dimension $D\ge4$.}

From now on, we put $D\ge4$. It holds: $ds^2$ is of constant
curvature iff
it is an Einstein space with vanishing Weyl tensor. In sect. IV.1
we review the
known cases, and in sect. IV.2 we give a general account of
conformal flatness.

\medskip

\section{IV.1. Dimension $D\ge4$. First part.}

Let $\d_\ab$ be the Kronecker tensor. The generalization of eqs.
(3.8), (3.9)
with $R=0$ reads
$$n=1,\qquad\qquad ds^2=dt^2\pm\d_\ab dx^\a
dx^\b\eqno(4.1)$$
and
$$n=2,\qquad\qquad ds^2=dt^2\pm(t^2\ dx^2+\d_\ab dx^\a
dx^\b)\eqno(4.2)$$
Generalizing (3.18), (3.19) with $R=0$ yields
$$n=1,\qquad\qquad ds^2=dt^2+t^2d\O_k^2\eqno(4.3)$$
and
$$n=1,\qquad\qquad ds^2=dt^2-t^2d\bar\O_k^2\eqno(4.4)$$
where $k=D-1$ and $d\O_k^2$ is the metric of the standard sphere
of dimension
$k$, while $d\bar\O_k^2$ denotes the corresponding space of
constant negative curvature.

For $R>0$ we get from (3.10) and (3.12)
$$n=1,\qquad\qquad ds^2=dt^2+\sin^2t\
d\O_k^2\eqno(4.5)$$
and
$$n=1,\qquad\qquad ds^2=dt^2-\sin^2t\ d\bar\O_k^2\eqno(4.6)$$
Eq.(4.6) is the \ads\st written as an open Friedman model for
$D=4$.

For $R<0$, we get from (3.13)
$$n=1,\qquad\qquad ds^2=dt^2\pm\et\d_\ab dx^\a dx^\b\eqno(4.7)$$
and from (3.20)-(3.23)
$$n=1,\qquad\qquad ds^2=dt^2+\sinh^2t\
d\O_k^2\eqno(4.8)$$
and 
$$n=1,\qquad\qquad ds^2=dt^2-\sinh^2t\ d\bar\O_k^2\eqno(4.9)$$
further  
$$n=1,\qquad\qquad ds^2=dt^2+\cosh^2t\ d\bar\O_k^2\eqno(4.10)$$
and 
$$n=1,\qquad\qquad ds^2=dt^2-\cosh^2t\ d\O_k^2\eqno(4.11)$$
The metric (4.11) represents the \des\st as a closed Friedman
model for $D=4$.
So, for every $D\ge 4$, we have found twelve representations with
$n=1$,
namely eqs. (4.1) and (4.3)-(4.11).

A further generalization of (4.3)-(4.4) is possible with $R=0$
and
$2\le k\le D-2$
$$n=2,\qquad\qquad ds^2=dt^2+t^2\ d\O_k^2+\d_\ab dx^\a
dx^\b\eqno(4.12)$$
and
 $$n=2,\qquad\qquad ds^2=dt^2-t^2\ d\bar\O_k^2-\d_\ab dx^\a
dx^\b\eqno(4.13)$$
Together with (4.2) this yields $2(D-2)$ \rep with $R=0$, $n=2$.

For $D=4$, the following \rep{}s are known, 
cf. eq. (1.2) above and ref. [4] for details.

\bigskip

$R>0$
$$
n=2,\qquad\qquad ds^2=dt^2+\sin^2t\ dx^2+\cos^2t\
d\O_2^2\eqno(4.14)$$
and
 $$n=2,\qquad\qquad ds^2=dt^2-\sin^2t\ dx^2-\cos^2t\
d\bar\O_2^2\eqno(4.15)$$

\bigskip

$R<0$
$$n=2,\qquad\qquad ds^2=dt^2+\sinh^2t\ dx^2+\cosh^2t\
d\bar\O_2^2\eqno(4.16)$$
and
$$n=2,\qquad\qquad ds^2=dt^2-\sinh^2t\ dx^2-\cosh^2t\
d\O_2^2\eqno(4.17)$$
further
 $$n=2,\qquad\qquad ds^2=dt^2+\cosh^2t\ dx^2+\sinh^2t\
d\O_2^2\eqno(4.18)$$
and 
$$n=2,\qquad\qquad ds^2=dt^2-\cosh^2t\ dx^2-\sinh^2t\
d\bar\O_2^2\eqno(4.19)$$
So, for $D=4$, we have already 10
 representations with $n=2$. By the
way, the
metric (4.17) is the same as (1.2).

\medskip

\section{IV.2. Conformal flatness.}

Spaces of \kc are conformally flat, and for $D\ge4$, conformal
flatness is
equivalent to the vanishing of the conformally invariant Weyl
tensor
$C^i_{\ jkl}$, defined as
$$ C_{ijkl} \ = \ R_{ijkl}-{1\over
D-2}(R_{ik}g_{jl}+R_{jl}g_{ik}
-R_{jk}g_{il}-R_{il}g_{jk})$$
 $$
+ \ 
{R\over(D-1)(D-2)}(g_{ik}g_{jl}-g_{jk}g_{il})\eqno(4.20)$$

Now, let $n=1$, i.e. $ds^2=dt^2\pm a^2(t)d\s^2$. According to
sect. III.3,
$d\s^2$ must be an Einstein space. But $ds^2$ is 
conformally flat, hence
$d\hat s^2=ds^2/a^2(t)$ is \cf too. After a time-rescaling, we
set
$$d\hat s^2=d\hat t^2\pm d\s^2$$
Inserting this metric into eq. (4.20), applying $\hat C_{ijkl}=0$
and the fact
that $d\s^2$ is an Einstein space, gives as a result that $d\s^2$
must be a
space of \kc. This proves that our classification for the case
$n=1$ given in
sect. IV.1 is already complete.

Let $n\ge2$ in the following. We want to show that every
$d\s_k^2$ is a space
of \kc. For $d_k=1$, this is trivial. For any $d_k\ge2$ we denote
$d_k$ by
$d$ and $D-d$ by $m$. Because of $n\ge2$ we have $m\ge2$. Let
$\dis=h_\ab
(x^\a)dx^\a dx^\b$ with $\a,\b=1,...,d$, and we write
$$ds^2=\pm\ak\dis+d\O^2$$
where $d\O^2$ is simply a description of the rest. Let
$d\hat s^2=\pm ds^2/\ak$, then we may write
$$d\hat s^2=h_\ab(x^\a)dx^\a dx^\b+h_\AB(x^A)dx^Adx^B$$
where $A,B=0,d+1,...,D-1$ ($h_\AB$ has unspecified signature).
$\dis$ has
\cur scalar $R_{(1)}$ and $d\hat\O^2=h_\AB dx^Adx^B$ \cur scalar
$R_{(2)}$.
Therefore, the metric $d\hat s^2$ has \cur scalar $\hat
R=R_{(1)}+R_{(2)}$.
Then we apply (4.20) and the fact that $C_{\a A\b B}$ vanishes.
This gives
$$R_\ab h_\AB+R_\AB h_\ab={\hat R\over D-1}h_\ab
h_\AB\eqno(4.21)$$

We calculate the trace of this equation by 
 multiplying with $h^\ab
h^\AB$.
Observing that
$d\ge2$ and $m\ge2$ this leads to
$${R_{(1)}\over d(d-1)}+{R_{(2)}\over m(m-1)}=0\eqno(4.22)$$
$R_{(1)}$ depends on $x^\a$ only, $R_{(2)}$ on $x^A$ only,
consequently both of
them are constant. By the way, for $m=d=D/2$ we get additionally
$\hat R=0$.
Knowing the constancy of $R_{(1)}$, we multiply eq. (4.21) by
$h^\AB$ and get
the result that $\dis$ is an Einstein space with constant \cur
scalar. In
the last step we apply $C_{\a\b\g\d}=0$ with eq. (4.20) to show
that $\dis$
is a space of constant curvature.

\medskip

\section{IV.3. Dimension $D\ge4$. Second part.}

According to the previous results, it remains to discuss the
cases with
$$n\ge2,\qquad\qquad ds^2=dt^2\pm\sum_{k=1}^n\ak\dis\eqno(4.23)$$
where each of the spaces $\dis$ represents a space of constant
curvature,
and the quotients $a_k(t)/a_j(t)$ are constant for $k=j$ only.
Let us start with the simplest case: every $d_k=1$, i.e.,
$n=D-1$,
and we have to check the positive signature case only
$$n\ge3,\qquad\qquad ds^2=dt^2+\sum_{k=1}^n\ak
(dx^k)^2\eqno(4.24)$$
We define $\a_k=\dot a_k/a_k$, $H=\sum\a_k$, $H_2=\sum\a_k^2$,
and
$\b_k=\a_k-H/n$. Consequently, $\sum\b_k=0$. The equation
$R_\AB=\l g_\AB$,
$\l$ = const, $A,B=0,...,n$ yields
$$\l+\dot H+H_2=0\eqno(4.25)$$
and 
$$\l+\dot\a_k+\a_kH=0\eqno(4.26)$$
Summing up the $n$ equations (4.26), we get
$$n\l+\dot H+H^2=0\eqno(4.27)$$
It holds: $a_k(t)/a_j(t)$ is constant iff $\b_k=\b_j$.

Let $B=\sum\b_k^2$, then it holds: the metric (4.23) has $n\ge2$
in the normal
form iff $B\ne0$. Proof: $B=0$ iff $\b_k=0$ for every $k$. But
equality of
all the $\b_k$ implies $\b_k=0$ because of $\sum\b_k=0$. q.e.d. 

Inserting
$H_2=B+H^2/n$ into eq. (4.25) we get
$$n\l+n\dot H+nB+H^2=0\eqno(4.28)$$
Eq. (4.26) can be written as
$$n\l+n\dot\b_k+\dot H+n\b_kH+H^2=0\eqno(4.29)$$
(Of course, summing up the $n$ equations (4.29) we recover
(4.27)).
The difference between (4.27) and (4.28) leads to
$$\dot H=-{n\, B\over n-1}\eqno (4.30)$$
Because of $B>0$ we have $\dot H<0$.

The difference between (4.29) and (4.27) reads
$$\dot\b_k+\b_k H=0\eqno(4.31)$$
This implies $\dot B=-2HB$ and with (4.30)
$$\left(\dot B\over B\right)^\cdot={2n\over n-1}B\eqno(4.32)$$
For $\l=0$, we get from (4.27) $H=1/t$, ($t>0$), i.e., with
(4.30),
$$B={n-1\over n\ t^2}\eqno(4.33)$$

With $H=1/t$, we can solve eq. (4.31), obtaining $\b_k=p_k/t$,
$p_k$ =
const. Together with eq. (4.33), we finally get
$$\sum p_k=0\qquad\qquad\sum p_k^2={n-1\over n}$$
Consequently, $\a_k=\b_k+1/(nt)=q_k/t$, with
$$\sum q_k=\sum q_k^2=1\eqno(4.34)$$
One can integrate these equations to $a_k(t)=t^{q_k}$ and eqs.
(4.34)
give the defining condition for the $d$-dimensional Kasner
solution.

Now we have to check under which conditions the Kasner solution
$$ds^2=dt^2+\sum_{k=1}^nt^{2q_k}(dx^k)^2\eqno(4.35)$$
is flat. (By the way, $n>1$ and (4.34) already imply $B>0$).
However, this
requires the calculation of the Weyl tensor.

\medskip

\section{IV.4. Calculating the Weyl tensor.}

Now we take the metric (4.24), assume that it is already an
Einstein space,
and calculate the Weyl tensor with the notations used in 
sect. IV.3.
In this section the sum conventions will not be used. It turns
out that the
non-trivial components of the Weyl tensor read
$$0=C^{0k}_{\ \ 0k}=-\dot\a_k-\a_k^2+c_1\eqno(4.36)$$
and
$$0=C^{jk}_{\ \ jk}=-\a_j\a_k+c_2\qquad
\qquad\qquad j\ne k\eqno(4.37)$$
where $c_1$ and $c_2$ are constants which vanish iff $ds^2$ has
vanishing
curvature scalar. Now we are prepared to continue our discussion.

\medskip

\section{IV.5. Dimension $D\ge 4$. Third part.}

Let $R=0$, i.e. $c_2=0$ in eq. (4.37). This means that at most
one of the
functions $\a_j$ may be different from zero. If all of them are
zero, we get
eq. (4.1), if one of them is non-zero, then we get (4.2), so no
new
representations are found here.

Let $R\ne0$, i.e. $c_2\ne0$ in eq. (4.37). Let $j,k,l$ be three
different indices
(which is possible because $n\ge3$). Then (4.37) implies
$c_2=\a_l\a_j=
\a_l\a_k\ne0$, hence $\a_l\ne0$, and consequently $\a_j=\a_k$.
Because of the
arbitrariness in the choice of $j$ and $k$, we conclude that all
the functions 
$\a_k$ coincide. So we have the case $n=1$ in the normal form
which has already
been discussed above. This concludes the discussion for the case
that all the
spaces $\dis$ are flat. By condition D) (see sect.II), we may now
assume that
$d\sigma^2_1$ has non-vanishing \cur and consequently $d_1\ge2$.
So we have to check under which circumstances (again, the index 1
will be
omitted)
$$ds^2=\pm a^2(t)d\s^2+dt^2\pm\sum_{k=2}^n\ak\dis\eqno(4.38)$$
represents a space of constant curvature. If $a(t)$ is constant,
this is
impossible. Hence, $\a=\dot a/a\ne0$. So, we have again the same
situation as
in sect. IV.2 and we may apply eq. (4.22) as follows: let
$$d\bar s^2={1\over
a^2(t)}\left[dt^2\pm\sum_{k=2}^n\ak\dis\right]\eqno(4.39)$$
then $d\bar s^2$ must be a space of non-vanishing constant
curvature.

Let $D=4$. Then necessarily $n=2$, and $\dis$ is 1-dimensional,
hence a flat
space and $d\s^2$ has to be a plane of constant curvature. The
results are
easily calculated and are given by eqs. (4.12)-(4.19), where we
have to
insert $k=2$ and $D=4$ in eqs. (4.12) and (4.13).

Higher dimensions can be discussed by induction over $D$: eq.
(4.39) has the
same structure (up to a redefinition of $t$), and therefore we
can deduce all
possible metrics, from which we can calculate the corresponding
metrics (4.38)
of dimension at least higher by two.

However, this is quite involved, so we prefer a different
approach. A more
careful inspection of eq. (4.38) leads to the observation that
eq. (4.37)
remains valid. Therefore, the discussion at the beginning of
sect. IV.5
still holds. Thus we can restrict to the case $n=2$ in the
following.
So we have to check:
\smallskip
$R=0$
$$ds^2=dt^2\pm(a^2(t)d\s^2+\d_\ab dx^\a dx^\b)\eqno(4.40)$$
with $\dot a=0$ and $d\s^2$ not flat.

\smallskip
$R\ne0$
$$ds^2=dt^2\pm(a^2(t)d\s^2+b^2(t)d\s^2_2)\eqno(4.41)$$
with $\dot a\ne0$, $\dot b\ne0$, and $d\s^2$ not flat.

For the case $R=0$, we get $a(t)=t$ and the metrics (4.12),
(4.13).
For the case $R\ne0$, we get the generalization of (4.14)-(4.19)
to
higher dimensions, which read
$$n=2,\qquad\qquad ds^2=dt^2+\sin^2t\ d\O_j^2+\cos^2t\
d\O_k^2\eqno(4.42)$$
and 
$$
n=2,\qquad\qquad ds^2=dt^2-\sin^2t\ d\bar\O_j^2-\cos^2t\
d\bar\O_k^2\eqno(4.43)$$
for $R>0$ and
$$n=2,\qquad\qquad ds^2=dt^2+\sinh^2t\ d\O_j^2+\cosh^2t\
d\bar\O_k^2\eqno(4.44)$$
and 
 $$n=2,\qquad\qquad ds^2=dt^2-\cosh^2t\ d\O_j^2-\sinh^2t\
d\bar\O_k^2\eqno(4.45)$$
for $R<0$, with $j+k=D-1$. In eq. (4.42), only $2 \le j \le k$
give new cases, these are $(D-4)/2$ for even $D$ and $(D-3)/2$
for odd $D$. The same result appears for eq (4.43). 
In eq. (4.44), all $j, \, k \ge 2$ give new cases, $D-4$ in sum,
as does eq. (4.45). 

\bigskip

\section{V. A more general ansatz.}

In this section, we consider a slightly 
different ansatz for the
metric of
a space of constant curvature, which is suitable for a 
direct
determination
of its general form from the solution of a simple system of
differential
equations. It is well-known, in fact, that  a \ms space is
uniquely
characterized by the property that
$$R_{ijkl}=\l(g_{ik}g_{jl}-g_{il}g_{jk})\eqno(5.1)$$
with a constant $\lambda$.
Adopting an appropriate ansatz for the metric, (5.1) gives a set
of differential
equations which determine the conditions under which the space is
of constant
curvature.

We take a metric of the form
$$ds^2=dt^2\pm\left[a^2(t)d\r^2+\sum_{i=1}^{D-2}
b_i^2(\r,t)(dx^i)^2\right]
\eqno(5.2)$$
where the metric functions depend on $t$ and one "spatial"
coordinate $\r$.
A more general form of the ansatz would require that $a=a(t,\r)$,
but our
choice is necessary in order to obtain a tractable set of
differential
equations from (5.1). Anyway, our choice is not restrictive since
the
metric (5.2) can be seen as a 
multi-warped product of a 2-dimensional
space,
spanned by the coordinates $t$ and $\r$, with a
($D-2$)-dimensional space,
and is well known that the metric of a 2-dimensional space
of constant curvature can
always be put
in the form $dt^2\pm a^2(t)d\r^2$ by a suitable choice of
coordinates.

For simplicity, in this section we consider the case of
Lorentzian signature.
The Euclidean case is easily obtained following the same
procedure.
Inserting the ansatz (5.2) into (5.1), one obtains the following
equations
$$\ddot a=\l a\eqno(5.3)$$
 $$\ddot b_i=\l b_i\eqno(5.4)$$
 $$\dot a\dot b_i-{b_i''\over a}=\l ab_i\eqno(5.5)$$
 $$\dot b_i\dot b_j-{b_i'b_j'\over a^2}=\l b_ib_j\eqno(5.6)$$
 $$a\dot b_i'-\dot ab_i'=0\eqno(5.7)$$
where a dot denotes a derivative with respect to $t$ and a prime
a derivative
with respect to $\r$. Eq. (5.7) implies that
$${d\over dt}\left({b_i'\over a}\right)=0\eqno(5.8)$$
Hence, either $b_i'=0$ or $b_i=a(t)g_i(\r)$.
We can now pass to discuss the solutions of (5.3)-(5.7). In the
following, we
tacitly assume that the \tran of section II can be applied to
change the form
of the solutions to their normal form.

\medskip

\section {V.1. $\l>0$.}

If $\l>0$, without loss of generality we can take $\l=1$ by 
condition B).
Eq. (5.3) then implies $a=Ae^t+Be^{-t}$ with $A$ and $B$
integration
constants. Moreover, if $b_i'=0$, eq. (5.4) implies that
$b_i=C_ie^{t}+
D_ie^{-t}$, with $C_i$ and $D_i$ integration constants.
Eqs. (5.5) and (5.6) then read
$$AD_i=-BC_i\qquad\qquad C_iD_j=-C_jD_i\eqno(5.9)$$
Only two distinct classes of solutions are avalaible: if either
$A=C_i=0$ or
$B=D_i=0$, one has for any dimension $D$ a solution of the
general form (4.7).
Further non-trivial solution are avalaible only in $D=3$, when
$D_1=-BC_1/A$,
or in $D=2$ and, modulo the transformations of sect. II they
reduce to
the form (3.14) in $D=3$ or (3.6), (3.7) in $D=2$.

If $b_i'\ne0$, instead, one has
$$b_i=A(t)g_i(\r)=(Ae^t+Be^{-t})g_i(\r)$$
which substituted in (5.5), (5.6) gives
$$g_i''=-4ABg_i\qquad\qquad g_i'g_j'=4ABg_ig_j\eqno(5.10)$$
One must distinguish the case $AB>0$ from $AB<0$.
If $AB>0$, $a(t)$ can be reduced to $\cosh t$, while, after a
rescaling of
$\r$, $g_i=F_i\sin\r+G_i\cos\r$ and (5.10) implies
$F_iF_j+G_iG_j=0$.
A solution to the last equation is available
only for $D\le4$, and can be put in the form\footnote{Here
and in the following,
lower dimensional solutions can be obtained by suppressing some
of the \coo
$x,y..$.}
$$ds^2=dt^2-\cosh^2t\ (d\r^2+\sin^2\r\ dx^2+\cos^2\r\
dy^2)\eqno(5.11)$$
which is of the type (4.11).

If $AB<0$, instead, $a(t)$ can be reduced to $\sinh t$, while
$g_i=F_ie^\r+G_ie^{-\r}$ and (5.10) implies $F_iG_j+G_iF_j=0$.
A solution to this equation valid in any
dimensions, is given by $F_i=0$ or $G_i=0$ and can be written
$$ds^2=dt^2-\sinh^2t\ (d\r^2+e^{2\r}\sum dx_k^2)\eqno(5.12)$$
Another solution is avalaible when $D\le4$,
if $F_2G_1=-F_1G_2$, which can be reduced to the form
$$ds^2=dt^2-\sinh^2t\ (d\r^2+\sinh^2\r\ dx^2+\cosh^2\r\
dy^2)\eqno(5.13)$$
Both solutions (5.11) and (5.12) are special cases of (4.9).
A last possibility remains to be investigated, namely the case in
which some
of the $b_i$ are independent of $\r$ and some are not. It is easy
to see that
in this case, eq. (5.6) is a consequence of (5.5) and hence one
gets two
new 5-dimensional solutions, and one in arbitrary dimensions, by
combining
(5.11)-(5.13) with (3.14)
$$ds^2=dt^2-\sinh^2t\ (d\r^2+e^{2\r}\sum dx_k^2)-\cosh^2t\
dy^2\eqno(5.14)$$
 $$ds^2=dt^2-\sinh^2t\ (d\r^2+\sinh^2\r\ dx^2+\cosh^2\r\
dy^2)-\cosh^2t\ dz^2\eqno(5.15)$$
 $$ds^2=dt^2-\cosh^2t\ (d\r^2+\sin^2\r\ dx^2+\cos^2\r\
dy^2)-\sinh^2t\ dz^2\eqno(5.16)$$
The solutions (5.14)-(5.16) are all special cases of (4.45).

\medskip

\section{V.2. $\l<0$.}

If $\l<0$, (5.3) yields $a=A\sin t+B\cos t$, with $A$ and $B$
integration constants.
If $b_i'=0$, eq. (5.4) is solved by $b_i=C_i\sin t+D_i\cos t$ and
eqs. (5.5)
and (5.6) imply
$$AC_i=-BD_i\qquad\qquad C_iC_j=-D_iD_j\eqno(5.17)$$
These relations admit solutions only for  $D=3$, for example
$B=C_1=0$ or
$A=D_1=0$. All these solutions are equivalent, via the
transformations of
section II, to (3.11).

If $b_i'\ne0$, instead, one has
$$b_i=A(t)g_i(\r)=(A\sin t+B\cos t)g_i(\r)$$
which substituted in (5.5), (5.6) give respectively,
$$g_i''=-(A^2+B^2)g_i\qquad\qquad
g_i'g_j'=-(A^2+B^2)g_ig_j\eqno(5.18)$$
These equations have solution $g_i=F_ie^\r+G_ie^{-\r}$, with
$F_iG_j+G_iF_j=0$.
Solutions to this equation exist up to 4 dimensions and take the
form 
$$ds^2=dt^2-\sin^2t\ (d\r^2+\sinh^2\r\ dx^2+\cosh^2\r\
dy^2)\eqno(5.19)$$
and hence are of the type (4.6).

Finally, in  the case in which some
of the $b_i$ are independent of $\r$ and some are not, it is easy
to see that
eq. (5.10) is always satisfied and hence one gets one new
5-dimensional
solution, which again is a special case of (4.43)
$$ds^2=dt^2-\sin^2t\ (d\r^2+\sinh^2\r\ dx^2+\cosh^2\r\
dy^2)+\cos^2t\ dz^2
\eqno(5.20)$$

\medskip

\section{V.3. $\l=0$.}

If $\l=0$, $a$ and $b$ must be linear in $t$. Proceeding as
before, one
obtains the following possibilities:
if $a=t$, $b_i$ is constant for any $i$ or can have the form
$b_1=t\sinh\r$, $b_2=t\cosh\r$. One then recovers the solutions
(4.2), or
$$ds^2=dt^2-t^2(d\r^2+\sinh\r^2\ dx^2+\cosh^2\r\
dy^2)\eqno(5.21)$$
which is equivalent to (3.19).
If $a$ is constant, one can have $b_1=\r$, $b_i$ = const, $i\ne1$
and hence
(4.13).

To conclude this section, we observe that no new solutions are
introduced
by the ansatz (5.2). However, we did not formally define which is
the generalization of condition F), and so in this sect. V,
the classification is not such a strict one as in sect. IV.
 This fact may lead to the conjecture that
even assuming
a more general dependence of the metric on the "spatial"
coordinates, no
new forms can be obtained for the solutions, in addition to those
listed in
the previous sections.

\bigskip

\section{VI. Discussion.}

Now we have finished the classification of the spaces of constant
curvature, and it is useful to sum up potential applications of
this.
First, consider any type  of cosmic no-hair theorem stating that
under
certain circumstances every solution of the Einstein field
equation is
\as a space of constant curvature. To find out regions where this
theorem
applies it is
necessary to know which spacetimes of a given class, e.g. the
multidimensional
warped products we considered here, is exactly a (generalized) de
Sitter
spacetime. We found unexpected forms on the one hand, and on the
other hand,
it came out that the cartesian product between a four-dimensional
\des\st
and one or more static internal spaces is never a space of \kc.

Example: metric (4.19) is a metric for a cosmological model of
Bianchi type III.
Usually, Bianchi type III consists of anisotropic models only,
and so (4.19)
may be used to test the range of validity of the no-hair theorem
within the
class of Bianchi type III models, which seems not to have been
done up to now.

This list may help identifying "new solutions" of Einstein's \fe
with or without
cosmological term in an arbitrary number of dimensions.
The most remarkable result of our paper seems to be that no
metric with more
than two factors is possible. If $n$ is the number of factors, we
have found
for every $D\ge2$ twelve \rep{}s with $n=1$ and
 $5D-10$ \rep{}s with $n=2$ if $D$ is even and 
 $5D-9$ \rep{}s with $n=2$ if $D$ is odd.

\bigskip

{\bf Acknowledgement}

HJS thanks the University of Cagliari for hospitality
and the DFG for financial support.

\beginref

\noindent 
\ref [1] U. Kasper, M. Rainer and A. Zhuk, Gen. Rel. Grav.
 {\bf 29}, 1123 (1997).

\noindent 
\ref [2] V. D. Ivashchuk and  V. N. Melnikov, 
Grav. and Cosmol. {\bf 2}, 211 (1996).

\noindent 
\ref [3] U. Bleyer, D.-E. Liebscher, H.-J. Schmidt and  A. Zhuk,
Wiss. Z. Univ. Jena {\bf 39}, 20 (1991).

\noindent 
\ref [4] H.-J. Schmidt,  Fortschr. Phys. {\bf 41}, 179 (1993).

\noindent 
\ref [5] J. Bi\v cak and J. Griffiths, Ann. Phys. (NY) {\bf
252}, 180 (1996).

\noindent 
\ref [6] M. Cadoni and S. Mignemi, Mod. Phys. Lett. A {\bf 10},
367 (1995).                      
\endref
\end{document}